\documentclass[a4paper,11pt]{article}
\pdfoutput=1 

\usepackage{jinstpub} 
\usepackage{algorithmic}
\usepackage{algorithm2e}   
\usepackage{graphicx}

\title{\boldmath Artificial Neural Network Algorithm based Skyrmion Material Design of Chiral Crystals}

\author[a]{B.U.V Prashanth,}
\author[1,b]{and Mohammed Riyaz Ahmed,\note{Corresponding author.}}

\affiliation[a]{Research Assistant}
\affiliation[b]{Associate Professor}
\affiliation[a,b]{School of Electronics and Communications Engineering, REVA University, Bengaluru-560064, India.}
\emailAdd{riyaz@reva.edu.in}

\abstract{The model presented in this research  predicts ideal chiral crystal and propose a new direction of designing chiral crystals. Skyrmions are topologically protected and structurally assymetric materials with an exotic spin composition. This work presents deep learning method for skyrmion material design of chiral crystals. This paper presents an approach to construct a probabilistic classifier and an Artificial Neural Network(ANN) from a true or false chirality dataset consisting  of chiral and  achiral compounds with 'A' and 'B' type elements. A quantitative predictor for accuracy of forming the chiral crystals is illustrated. The feasibility of ANN method is tested in a comprehensive manner by comparing with  probalistic classifier method. Throughout this manuscript we present deep learnig algorithm design with modelling and simulations of materials. This research work elucidated paves a way to develop sophisticated software tool to make an indicator of crystal design.}


\proceeding{N$^{\text{th}}$ Workshop on X\\
  when\\
  where}
\begin{document}
	\maketitle
	\flushbottom
	\section{Introduction:}
	In year 2018, Eri Shimono et al. developed the machine learning based  statistical method to predict  the chemical categories ideally suitable for creating chiral molecules\cite{Shimono2018}. The model elucidated in current research further enhances the above work with deep learning approach employing probabilistic classifiers and artificial neural networks. The properties of heterochiral structures of skyrmions have been reported by several research groups\cite{PhysRevB.95.144401,Li2018,Bourianoff2018}. Current day artificial intelligence combined with deep learning based statistical techniques have gained wide a popularity for material design. In this novel paradigm the deep learning methods make predictions for novel un-characterized materials and invent completely new materials\cite{Granda2014}. The artificial neural networks have the ability to express complex function mappings without the knowledge of  structural features and they form to be an effective estimation technique in the analysis of non linear behaviours. Input and output relationship of the system can be predicted through artificial neural networks irrespective of understanding of detailed mechanisms involved \cite{Kaubruegger2018}. At present there is no fixed method to predict the accuracy of forming chiral crystals and relies on  several logical models and further the  coexistence of chiral geometry in a single crystal when various atoms are mixed is difficult to predict\cite{Hsieh2009}. There is a urgent need of a statistical technique  which can predict between two objects. Naive bias probalistic classifier is one of the statistical technique which can exhibit best accuracy and adaptability in prediction of chirality. It is great potential to obtain dataset from large crystallographic databases such as ICSD\cite{Groom2016}  and apply artificial neural networks and probabilstic classifiers for drawing the insights from huge quantity of data and  analyzing the dataset.
\section{Data Collection and Modeling Scheme}
The database was built with properties of chemical elements including materials which interact and move as particle of topological spin texture\cite{Leliaert2018}. Chiral compounds and achiral AxBy type compounds with A Group and B Group elements are extracted with true or false chirality dataset. The selection of features is carried out with according to pre-defined scientific understanding\cite{Nord1998}. In this research work three variables such as group numbers of 'A' and 'B', and true or false chiral are selected as features and applied to  probalistic classifier. The following algorithm describes this classification process in detail. Here the input data for probabilistic classification is based on group number of 'A','B' and true or false chiral. Further probabilistic computation is performed to find out the chance of variation among true or false chiral in the input dataset. The probabilistic classifier selected in this research is naive Bayes classifier. 
\newline
\newline 
\begin{algorithm}[H]
	\SetAlgoLined
	\KwResult{Performance of True or False Chiral Dataset}
	Model initialization and Importing dataset\;
	\While{Read the CSV file}{
		Convert categorical variable to numeric\;
		Split dataset in training and test datasets\;
		Instantiate the classifier\;
		\eIf{Guassian Naive Bayes}{
			Select Ax and By as used features\;
			Select chirality as target feature\;
		}{  \textit{Compute:}
			\newline mean and standard deviation of chirality and achirality\;
			number of mislabeled points out of total points\;
		}
		Calculate the performance\;
	}
	\caption{naive Bayes Probabilistic classifier }
\end{algorithm}
\section{Artificial Neural Network Model}
An artificial neural network model was built based on dataset and selected features with three layers as 3*5*2 structure with input,hidden and output layers. This ANN model is compared with the  naive Bayes probabilistic classifier which is as elucidated in the above section. Using tensor flow framework the ANN model was built and using Scikit-learn package the probabilistic classifier was achived in Python environment. The following algorithm illustrates the detailed steps of ANN model implementation. 
\begin{algorithm}
	\SetAlgoLined
	\KwResult{True or False chirality}
	Import dataset \;
	Define Nodes and Learning rate\;
	Training epochs and weights\;
	Weight optimization is Gradient Descent
	\While{Target = Chirality}{
		\textit{Assume:} 
		L2 loss function\;
		Sigmoid Activation\;
		No bias terms\; 
		\eIf{normalize features}{
			Create cross-validation folds\;
			Train/evaluate the model on each fold\;
			Collect training and test data from folds\;
			Build neural network classifier model and train\;
			Make predictions for training and test data\;
		}{  \textit{Compute:}
			training/test \%error\;	
		}
		Visualize classification accuracy\;
		Plot the Histogram Spectra\;
	}
	\caption{Illustration Artificial Neural Network Model}
\end{algorithm}
The optimization is based on gradient descent algorithm and is very useful in process of manufacturing a semiconductor.
The figure 1 illustrates the diagram of artificial neural network with the accurate implementation steps is as described in algorithm2.
The process factors such as Ax, By, chirality True/False for respective compounds\cite{PhysRevB.99.104409} are applied as input data and are  utilized in each run of Artificial Neural Network(ANN) model. The performance of the ANN model is measured by RMSE and to determine performance of fitted line describe the data by measuring  R$^2$ score. The following table 1 summarizes the RMSE and R$^2$ values of chirality and achirality modelling with training and testing data. The obtained RMSE and R$^2$ values of training data are comparable to the obtained RMSE and R$^2$ values of testing data.
\begin{table}
	\caption{Neural Network Parameters with Training and Testing Data}
	\label{tab:1}       
	\centering
	\begin{tabular}{p{2cm}p{3cm}p{3cm}p{0cm}}
		\hline
		Parameter&training data&testing data\\
		&RMSE \hspace{0.5cm}R$^2$&RMSE\hspace{0.5cm}R$^2$\\
		\hline
		Chirality&0.04439\hspace{0.5cm}91.63&0.00262\hspace{0.5cm}95.92\\
		A-chirality&0.4906\hspace{0.7cm}97.03&0.189\hspace{0.9cm}94.12\\
		\hline
	\end{tabular}
\end{table}
\begin{figure}[htb]
	\centering
	\includegraphics[scale=0.60]{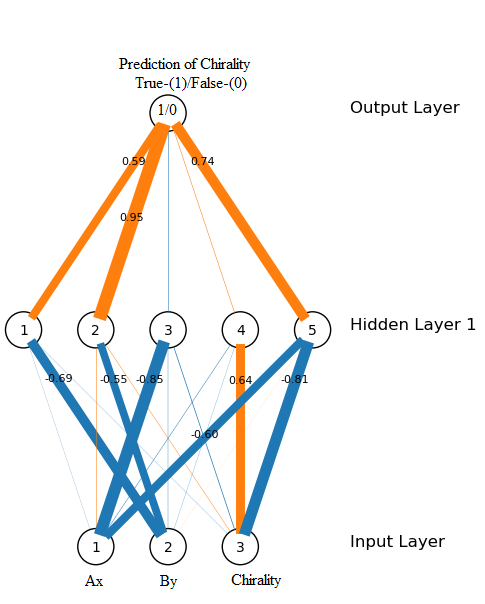}
	\caption{ANN model Implementation}
	\label{fig:1}       
\end{figure}
\section{Results}
The probabilistic classifier as discussed in section 2 is developed to predict the true or false chirality in the given data set.
The results are obtained in the form of mean and standard deviation of true or false chirality along with number of mislabeled points out of a total points and performance metrics as depicted in table 1.
To test the feasibility of ANN method in a comprehensive way we compare it with  naive bayes classifier method. 
\begin{figure}[htb]
	\centering
	\includegraphics[scale=0.60]{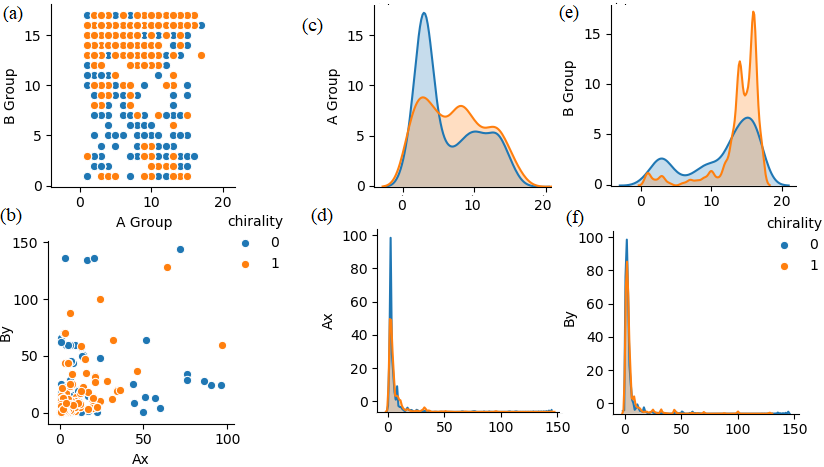}
	\caption{(a)chirality separation in A Group vs B Group elements and (b)chirality saperation in Ax vs By with True or False chirality, (c),(d),(e) and (f) Histogram Spectra of Chirality True/False dataset obtained from Probabilistic Classifier}
	\label{fig:2}       
\end{figure}
The Fig.2(a) illustrates the chirality separation characteristics obtained from probabilistic classifier and the Fig.2(b) depicts chirality saperation in Ax vs By with True or False chirality . Fig.2(c),(d),(e) and (f) describe the comparisions of 'A' Group versus 'B' Group and 'Ax' versus 'By' histogram spectra predictions.
\begin{table}
	\caption{Performance of Naive Bayes Classifier}
	\label{tab:2}       
	\begin{tabular}{p{6cm}p{2.5cm}p{2.5cm}p{2cm}}
		\hline
		Parameter&Chirality&A-Chirality&Value\\
		\hline
		Standard Deviation&8.92&9.09&-\\
		Mean Deviation&4.6&3.72&-\\
		Number of Mislabeled points&-&-&328 out of 843\\
		Performance&-&-&61.09\%\\
		\noalign{\smallskip}\hline\noalign{\smallskip}
	\end{tabular}
\end{table} 
The True or False chirality can be predicted from ANN model after 100000 epochs trains. A histogram spectra of classification accuracy obtained after training the ANN is as depicted in Fig.3 which describes the best performance in terms of accuracy. This graph illustrates the comparison of true/false chiral data labels from the evaluation set and along with the labels predicted by the trained network with only 3.6\% error. The neural network modelling results depict the comparisions of chirality and A-chirality\cite{Xu2006} are illustrated and here the line indicates the perfect fit between neural network and the experimental data.
\begin{figure}[htb]
	\centering
	\includegraphics[scale=0.60]{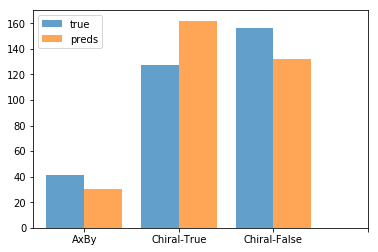}
	\caption{Histogram of classification accuracy}
	\label{fig:3}       
\end{figure}
The assumption made is the experimental errors are normally and independently distributed. A satisfying compability between training and testing values is based on modelling results of the neural network model. 
\begin{figure}[htb]
	\centering
	\includegraphics[scale=0.50]{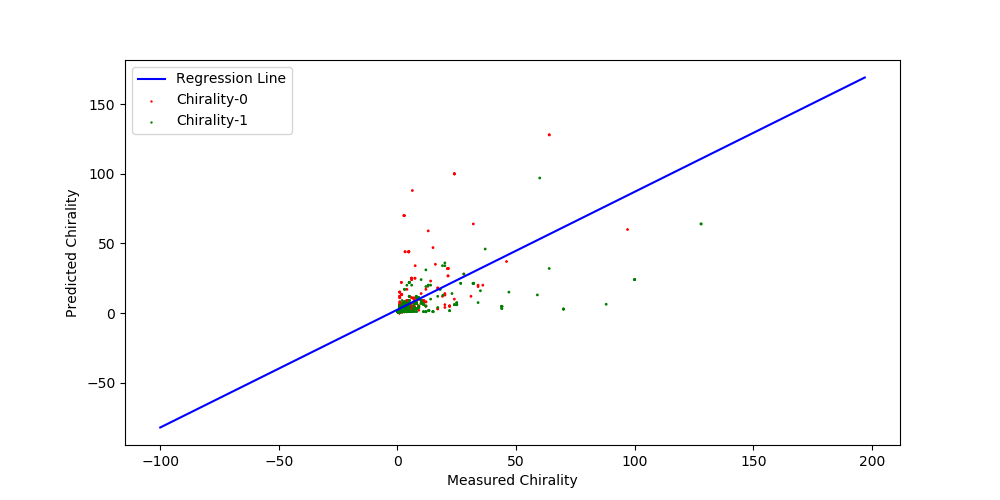}
	\caption{Neural Network Modelling Results for Chirality}
	\label{fig:4}       
\end{figure}
The comparision between chirality and A-chirality is plotted in figures 4 and figures 5. As the number of samples increases the accuracy of predictions will also increase, and further the test systems, contents of the crystal material compounds and the AxBy distribution differ in a wide range\cite{Woo2016}. Therefore for the purpose of comparision, the data is applied to training a neural network and perform  linear regression. The RMSE values of testing data were least  or comparable  to those of training data and for testing and training data of R$^2$ score values are over 90\%. The obtained modelling results of linear regression analysis from neural network training  it is verified that  there is a difference between chirality and a-chirality in True or False chirality dataset. 
\begin{figure}[htb]
	\centering
	\includegraphics[scale=0.50]{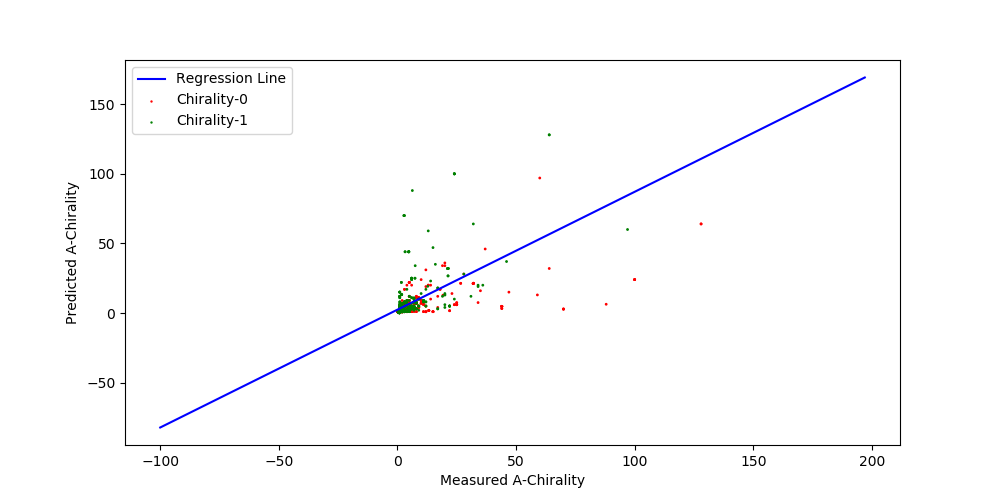}
	\caption{Neural Network Modelling Results for A-Chirality}
	\label{fig:5}       
\end{figure}
\section{Conclusions}
This research work utilized probabilistic classifier and ANN methods to classify the chirality True/False dataset in skyrmion material design of chiral crystals. Accceptable prediction outcome is achieved by ANN model. The ANN model method is framed to assist researchers in the domain areas of material science. In future this model could become an indicator of crystal design.

\acknowledgments
The authors would like to thank REVA University, Bengaluru, for providing all the necessary facilities to carry out this research work.
\bibliographystyle{JHEP}
\bibliography{ref}
	
\end{document}